# Allometric scaling in-vitro

Arti Ahluwalia

Department of Information Engineering and Research Center E.Piaggio, University of Pisa, Pisa, Italy.

*Correspondence to: arti.ahluwalia@unipi.it

ABSTRACT

The quarter power allometric scaling of mammalian metabolic rate is largely regarded as a universal law of biology. However, it is well known that cell cultures do not obey this law. The current thinking is that were in-vitro cultures to obey quarter power scaling, they would have more predictive power and could for instance provide a viable substitute for animals in research.

About two decades ago, West and coworkers established a model which predicts that metabolic rate follows a quarter power relationship with the mass of an organism, based on the premise that tissues are supplied nutrients through a fractal distribution network. This paper outlines a theoretical and computational framework for establishing quarter power scaling in-vitro, starting where fractal distribution ends. Allometric scaling in non-vascular spherical tissue constructs was assessed using models of Michaelis Menten oxygen consumption and diffusion. The results demonstrate that physiological scaling is maintained when about 5 to 60% of the construct is exposed to oxygen concentrations less than the Michaelis Menten constant. In these conditions the Thiele modulus is between 8 and 80, with a significant concentration gradient in the sphere. The results have important implications for the design of downscaled in-vitro systems with physiological relevance.

INTRODUCTION

Allometric scaling laws, which correlate the mass of organisms with physiological parameters through an exponent "b", have been explored by scientists for well over a century. Probably the best known allometric relationship is the ¾ power law describing the mass (M) and energy consumption or metabolic rate (MR) correlation (Eq 1). The ¾ power law and its related mass specific metabolism or cellular metabolic rate (CMR), which scales with an exponent of b=–¼, have been the subject of numerous publications [1–5]. Although the reason behind allometric scaling of metabolic rate is still not clearly understood, the remarkable consistency of the so-called "quarter power laws" for metabolic rates and other metabolism related parameters over several orders of magnitude of mass have led West and Brown, widely considered as the current gurus of allometry, to deem them universal laws in biology [4]. In essence, biological organisms are assembled according to the same basic design principles and using the same building blocks (mainly water) such that self-similarity is preserved across all scales [6].

The ¾ power allometric scaling law for MR is:

$$MR = a_b M^{3/4} \qquad (1)$$

Where, $a_b$ is a constant for all mammals, M is body mass in kg and MR is the whole body metabolic rate, here expressed in moles of oxygen consumed/s. Almost two decades ago West, Brown & Enquist used the fact that many organisms have fractal-like networks for resource transport to predict a value of ¾ for b.

Given that cell density (ρ in #cells/m³) in mammals is constant and thus mass invariant, and that the density of biological organisms is close to that of water (Ω=1000 kg/m³), the metabolic rate per cell, or CMR in moles of oxygen/(cell.s) can be expressed as:

$$CMR = \frac{MR}{\#cells} = \frac{a_b}{\rho} \Omega M^{-1/4} = a_c M^{-1/4} \qquad (2)$$

A log-log graph of CMR against mass gives a straight line with a slope of b=- ¼ (Figure 1), indicating that the oxygen consumption rate per cell increases in an organism as its body mass decreases.

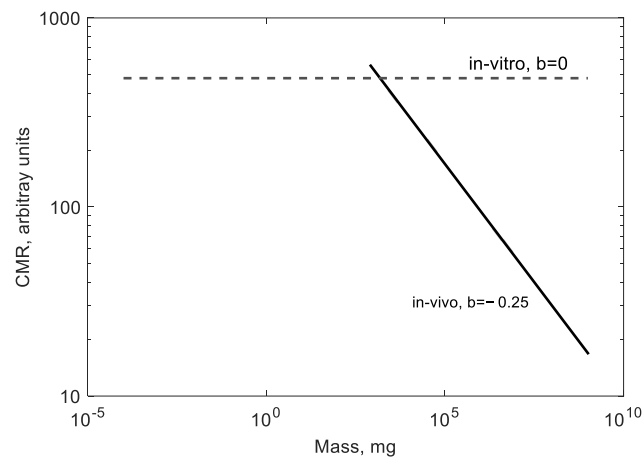

*Figure 1: CMR as a function of the mass of an organism in-vivo and in-vitro on a log scale plot. The allometric exponent b is the slope of the line. The cellular metabolic rate increases when cells are removed from the in-vivo context and cultured in the laboratory. Theoretical considerations (see text) show that this behaviour is due to high oxygen concentrations and the lack of oxygen gradients in-vitro. Figure adapted from [7].*

It is well known that different tissues and cell types consume oxygen at different rates according to their metabolic requirements [8] and the CMR in Eq. 2 therefore represents an average metabolic rate per cell in an organism. The CMR versus mass relationship has been explored by West and co-workers [7]. By examining CMR data from cells in culture and comparing them with whole animal CMR values (derived using equations 1 and 2) they demonstrated that metabolic scaling is not conserved in-vitro. For all mammals, individual cells in-vitro consume oxygen at a faster rate than in-vivo. Moreover, the CMR of cultured mammalian cells converges to an approximately constant value independent of the mass of the animal of origin. They hypothesized that the number of mitochondria per cell from any mammal settles to a constant maximal value after several

A.Ahluwalia, Allometric scaling in-vitro, August 2016

generations in culture. Thus the log-log slope of in-vitro CMR versus the mass of the animal from which they derive is near zero (Fig 1). A recent analysis of in-vitro and in-vivo oxygen consumption rates suggests that when hepatocytes are freshly isolated their CMR follows somewhat less than – ¼ power scaling [9]. The author (Glazier) concludes that metabolic rate and its scaling with mass is not only determined by energetic or physical constraints but also by systemic regulation.

The application of allometry to the design of in-vitro systems was first proposed by Vozzi et al. in 2009 [10]. Allometric scaling was used to engineer a "physiologically relevant" on-a-plate multi-organ with the objective of somehow extrapolating the results to human physiology. Since then several researchers have discussed the use of allometry for scaling down human body parameters to on-chip or on-plate devices, although very few have actually implemented systems based on allometric scaling [11,12]. Ucciferri et al. [13] show that scaling cell numbers in a 2 compartment hepatocyte and endothelial cell model can better mirror human glucose metabolism than scaling metabolic rate and surface area. However, despite current efforts to design organ and body-on-chip systems, in-vitro models of biological tissue are considered functionally inferior to whole organisms and their translational potential is still limited. In a seminal paper, Moraes et al. [14] suggested that in-vitro or on-chip systems should be designed on the principle of conserving "metabolically-supported functional scaling" so as to maintain in-vivo cellular metabolic rates when down-sized. Based on this principle, physiologically meaningful in-vitro on-chip or on plate multi-organ systems should follow the quarter power allometric laws expressed in Eqs. 1 and 2 as the size of devices is reduced to the microscale. One of the strategies proposed was to somehow modulate oxygen supply to cells to control their metabolic consumption.

Here the metabolic response of in-vitro constructs was characterized to determine a working window in which engineered tissues maintain biological metabolic scaling in the absence of vascularization As cellular oxygen consumption is regulated by Michaelis Menten (MM) reaction kinetics [15], the average MR and CMR of cell-filled spheroids were determined using computational mass transfer models coupling the MM reaction of oxygen and its diffusion through the construct.

THEORY

Assuming spherical symmetry, the reaction-diffusion equation in spherical coordinates for Michaelis Menten mediated oxygen consumption is:

$$\frac{\partial c}{\partial t} = \frac{D}{r^2}\frac{\partial}{\partial r}\left(r^2\frac{\partial c}{\partial r}\right) - \frac{V_{max}c}{k_m + c} \qquad (3)$$

Where $c$ is the oxygen concentration in moles/m³, D is the diffusion constant of oxygen in water (m²/s) at body temperature, $V_{max}$ is the maximum oxygen consumption rate per unit volume (moles/(m³.s)) and $k_m$ is the MM constant in moles/m³. Both $k_m$ and $V_{max}$ are rather difficult to measure, particularly in-vivo. Typically, $V_{max}$ is derived indirectly from measurements of flow rates and arterial and venous oxygen concentrations in organs in-vivo. In-vitro it is expressed as the product ρ×CMR [15]. The constant $k_m$ is estimated by fitting oxygen consumption data versus ambient oxygen concentration curves [16].

Both $k_m$ and CMR estimated from isolated tissues and cells have a narrow range of values (within one order of magnitude) in the literature, even amongst different species [7,15,16]. On the other hand, $V_{max}$ can be thought of as the product of enzyme affinities in mitochondria - which are approximately constant across several species [17,18] - and the overall density of mitochondria in a tissue. According to West et al.'s extrapolations, in in-vitro conditions the mitochondrial density in cells reaches an upper limiting value of ≈ 300/cell [7]. Therefore, at least in in-vitro monolayer cultures, $k_m$ and the limiting CMR, can be assumed as material constants related to mitochondrial enzymes. Hence, without any loss of generality and in the light of the fairly narrow range of reported oxygen consumption rates in-vitro , the values of $k_m$ and limiting CMR used here are from a previous study on scaling in hepatocyte cultures [19]: respectively 7.39×10$^{-3}$ moles/m$^3$ and 4.8×10$^{-17}$ moles/(cell.s).

Considering a representative spherical tissue or cell construct with radius $R$ (in m), the overall MR is the total inward flux at its surface multiplied by the total surface area.

$$MR = D \frac{dc}{dr}\bigg|_R 4\pi R^2 \qquad (4)$$

Given that a concentration gradient forms within the sphere, each cell will have a different consumption rate according to the oxygen concentration it perceives. The average CMR is then simply the MR divided by the total number of cells in the sphere.

$$CMR = D \frac{dc}{dr}\bigg|_R \frac{3}{\rho R} \qquad (5)$$

As outlined in the Appendix, when the oxygen concentration is over an order of magnitude greater than $k_m$ throughout the sphere (i.e. $c \gg k_m$), the reaction rate is zero order and the CMR is independent of $R$ and the same for all cells.

$$CMR = \frac{V_{max}}{\rho} \qquad (6)$$

This very simply explains why CMR estimated in-vivo is lower than that measured in-vitro. Even in the absence of an increase in mitochondrial density in-vitro, when isolated from an organism cells are usually plated in monolayers and all perceive the same oxygen concentration. As $k_m$ is typically 10 to 100 times less than the dissolved oxygen concentration in water at 37°C (0.2 moles/m$^3$, 0.2 mM or 150 mmHg)[15,16], the cell oxygen consumption rate is zero order. Conversely, within an organism oxygen supply is limited and its concentration depends on the distance between cells and their arterial blood supply. Therefore, for a given organism, the average CMR in-vivo is lower than the CMR in monolayer cultures (Figure 1).

On the other hand, when $c \ll k_m$, the rate is first order and the CMR depends on the ratio of the reaction rate to the diffusion rate (or the Thiele modulus, $\phi^2$). As shown in the Appendix, a quarter power metabolic scaling holds when the Thiele modulus, $\phi^2$ =24.808.

COMPUTATIONAL MODELS

As Eq. 3 cannot be solved analytically, to determine a working window in which metabolic scaling follows the – ¼ power allometric law for CMR, an array of 3D (three-dimensional) cell-filled spheres was modelled using the mass transfer module in COMSOL Multiphysics (version 3.5a COMSOL AB, Stockholm, Sweden). The constants and conditions used are listed in Table 1.

| Parameter | Symbol | Range | Notes | Ref |
|---|---|---|---|---|
| Construct radius | R | 31 – 5000 μm | From cell monolayers to 5 mm thick spheroids or tissue constructs. | [19] |
| Initial and boundary oxygen concentration | $c_o$ | 0.2 - 0.03 moles/m$^3$ or mM | Boundary conditions at the surface of the construct: implies media is well-mixed and continuously renewed at the surface. | [20] |
| Single cell CMR | | 4.8×10$^{-17}$ moles/(cell.s) | Typical value for hepatocytes in-vitro. | [15,19] |
| Cell density | ρ | 5.14×10$^{14}$ - 5.14×10$^{12}$ cells/m$^3$ | From physiological cell density to typical in-vitro density (5 million cells/mL). | [21] |
| *Michaeles Menten constant* | $k_m$ | 7.39×10$^{-3}$ moles/m$^3$ | Typical value for hepatocytes. | [19] |
| Oxygen diffusion constant | D | 3×10$^{-9}$ m$^2$/s | Diffusion constant in water at 37°C. | [19] |
| Reaction rate | | $-\dfrac{V_{max}c}{k_m+c}$ | Michaelis Menten consumption. | |
| Stationary conditions | | $\dfrac{\partial c}{\partial t}=0$ | | |

*Table 1: Parameters and boundary conditions input to the computational models.*

The models were solved in stationary conditions using the UMFPACK direct solver. The computational grid (or mesh) was generated using the COMSOL predefined "Fine" mesh size for the 3D spheres. Once the solutions were obtained, the surface and domain integration functions were used to calculate the total inward flux at the boundaries, average CMR and the volume of the sphere operating at $c<k_m$. Concentration gradients were determined from domain cross section plots and the Thiele modulus was computed from [22]:

$$\phi^2 = \dfrac{R^2}{D}\dfrac{V_{max}}{(k_m+c_o)} \qquad (7)$$

Finally, the data were imported into Matlab for plotting and curve fitting (Matlab R2015a, The Mathworks, USA. Curve Fitting Toolbox).

Figure 2 reports the resulting log scale graph for CMR versus mass for constructs exposed to a surface oxygen concentration ($c_o$) of 0.2 moles/m$^3$ with a range of cell densities from the physiological value of 5.14×10$^{14}$ to 5.14×10$^{12}$ cells/m$^3$. The latter corresponds to about 5 million cells/mL, which is considered a fairly high in-vitro density, often used for encapsulating cells in alginate-based microspheres [23]. As predicted by the analytical zero order model (see Appendix), at low values of construct mass, the CMR does not change but tends to a constant limiting value for all spheres. However, at a certain critical mass the slope becomes negative, reaching a steady value of -1/3 for the largest spheres with high cell density. The quarter power law holds in the region where the slope is around – ¼. Thus the window for physiological metabolic scaling was estimated by least squares fitting the linear region of the log-log plots, such that a slope of -0.25 was within the 95% confidence limits of the fitted line, with an $R^2$ > 0.99.

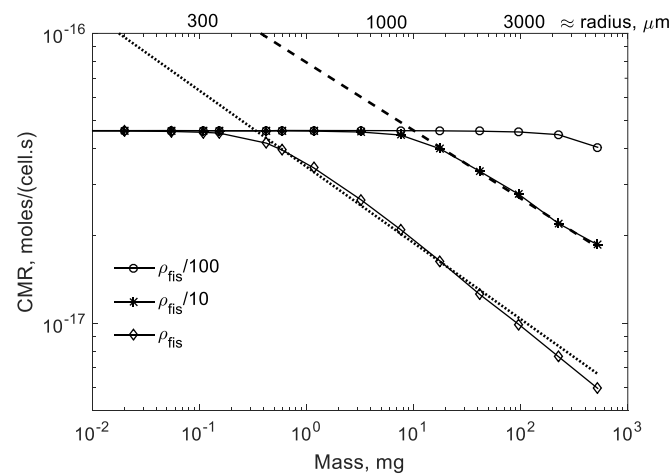

*Figure 2: Computed CMR against construct mass for different cell densities with ambient oxygen=0.2 moles/m$^3$.* Diamonds (◊): physiological density $\rho_{fis}$= 5.14×10$^{14}$ cells/m$^3$, asterisks (*): $\rho_{fis}$/10, circles (o), $\rho_{fis}$/100. Dotted line: least squares fit with slope= -0.2631±0.0222, $R^2$=0.9946. Dashed line: least squares fit with slope= -0.2348±0.0395, $R^2$=0.9970.

For the lowest cell density, only in the largest construct is there a notable decrease in CMR. On the other hand, the two higher cell densities have a narrow window of characteristic dimensions in which the slope closely approximates the – ¼ necessary for physiological metabolic scaling. The spheres with physiological cell density ($\rho_{fis}$) that fall within the quarter power window are surprisingly large considering that the maximum intercapillary distance in tissues is around 200 µm [24]. There is now a large body of evidence showing that tissues in-vivo thrive at low oxygen levels, certainly much lower than those routinely used in-vitro [20,25]. To examine the effect of ambient $O_2$ concentration on the CMR, the same reaction and diffusion equation was solved for a boundary concentration of 0.0733 moles/m$^3$, corresponding to 55 mmHg (rather than 150 mmHg) of $O_2$ partial pressure, which is at the higher end of measured $O_2$ levels in human livers [20]. The results are plotted in Figure 3A and show how the physiological scaling region is left-shifted towards smaller constructs (down to radii of 200 to 300 µm) as oxygen concentrations are reduced to in-vivo normoxic levels. This intriguingly suggests that intercapillary distances in-vivo are gauged so as to just tip tissues from zero to quarter power scaling and could be used as a design guideline for in-vitro constructs.

The effect of reducing ambient oxygen in-vitro to compensate for the lower cell densities usually employed in cell culture was assessed. Figure 3B illustrates how a reduction in surface oxygen levels again shifts the window for physiological metabolic scaling towards smaller spheres.

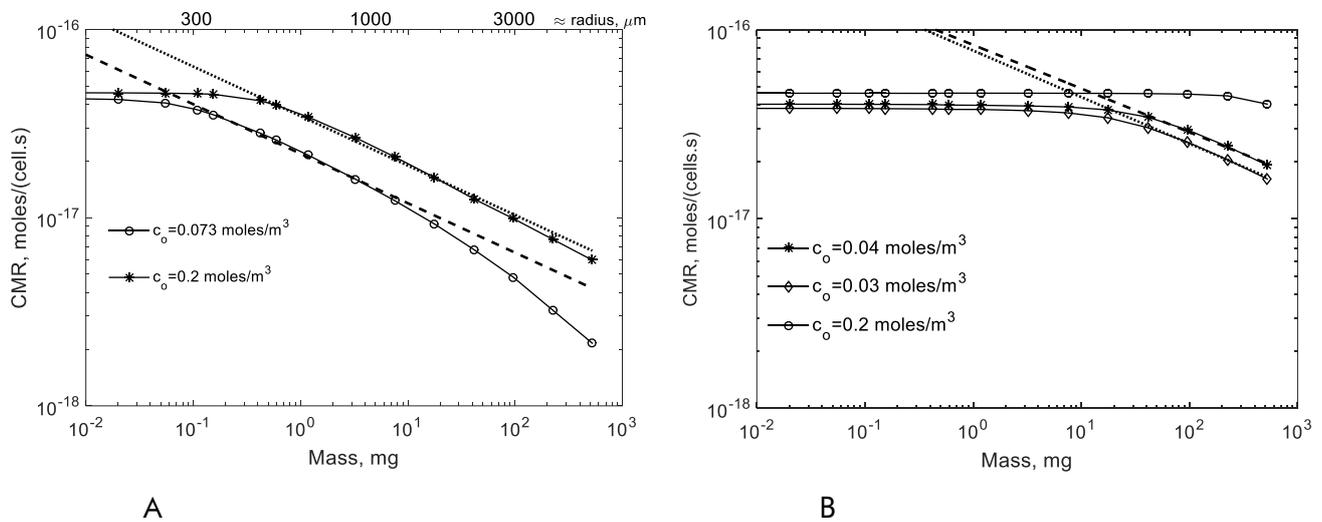

*Figure 3: CMR versus construct mass for different ambient oxygen levels. A) Physiological cell density ($\rho_{fis}$). Asterisks (\*): dissolved oxygen in atmospheric conditions, 150 mmHg $O_2$ (0.2 moles/m³), circles (o) 55 mmHg $O_2$ (0.073 moles/m³). Dashed line: least squares fit with slope= -0.2631±0.0222, $R^2$=0.9946. Dotted line: least squares fit with slope= -0.2630±0.0239, $R^2$=0.9938. B) In-vitro cell density $\rho_{fis}$/100. Circles (o) dissolved oxygen in atmospheric conditions, 150 mmHg $O_2$ (0.2 moles/m³), asterisks (\*) 30 mmHg $O_2$ (0.04 moles/m³) diamonds (◊) , 22.5 mmHg $O_2$ (0.03 moles/m³) . Dashed line: least squares fit with slope=-0.2309 ±0.0579, $R^2$ =0.9932, dotted line least squares fit with slope= -0.246 ±0.0478, $R^2$= 0.9959.*

Finally, to establish a set of generic quarter power scaling determinants, the fraction of construct volume with oxygen concentrations below $k_m$ and the Thiele modulus ($\phi^2$) were plotted as a function of radius (Figures 4A and B). Essentially, the range of masses identified in Figures 2 and 3 as falling within the window of physiological metabolic scaling correspond to spheres in which around 5 to 60% of the volume is at concentrations less than $k_m$ and $\phi^2$ lies between ≈8 and 80. In these conditions there is a notable oxygen gradient within the construct (Figure 4C).

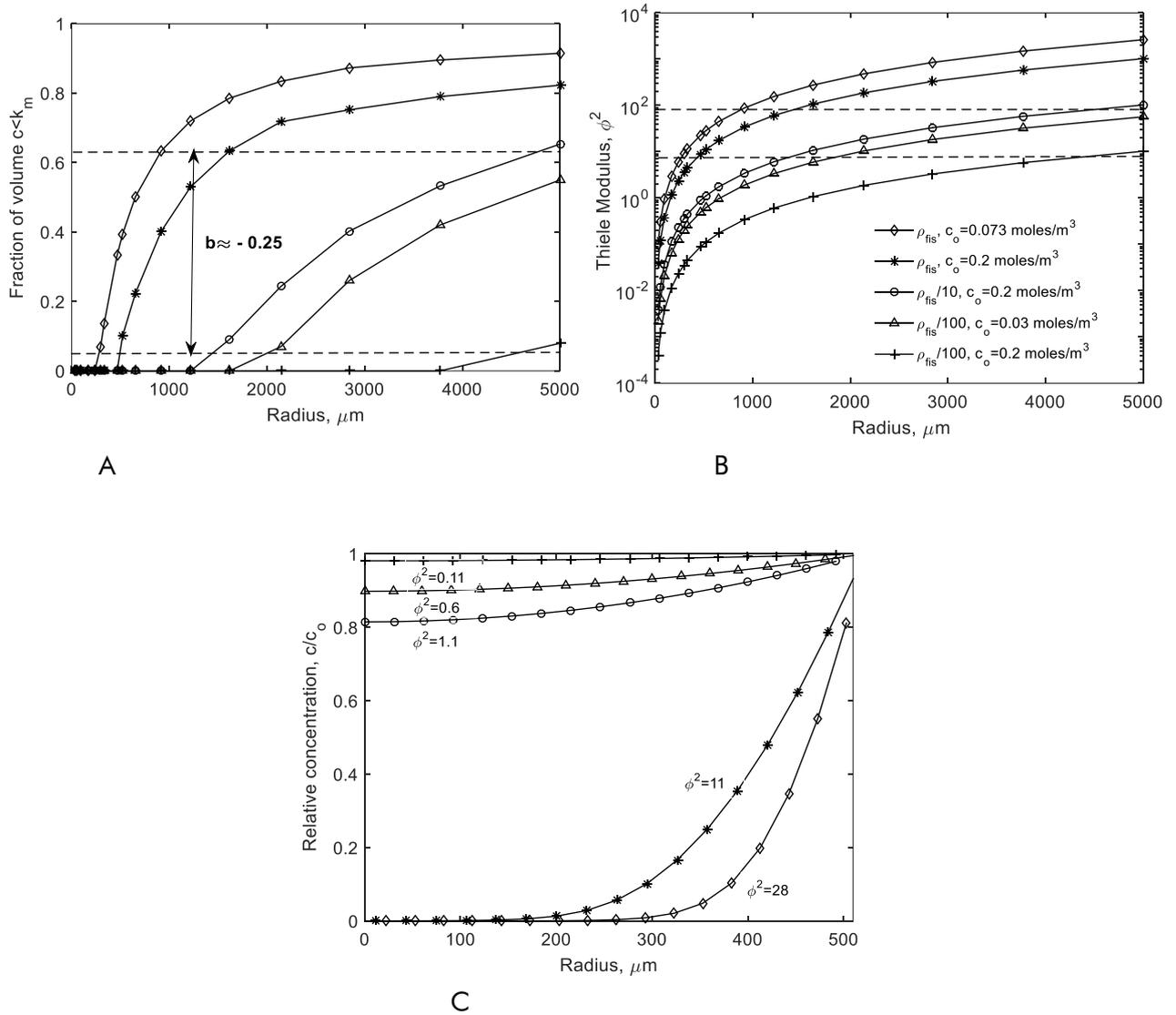

*Figure 4: Identifying a working window for physiological scaling. A) Fraction of construct volume at $c<k_m$ versus construct radius showing the working window for quarter power scaling between the dashed lines. B) Thiele modulus (from Eq. 7) versus radius. The quarter power window lies between the dashed lines. C) Concentration gradient in 521 μm radius ($\approx$ 0.6 mg) spheres with different values of cell density and ambient oxygen showing $\phi^2$ for each case. For this construct size, quarter power +scaling was observed only for $\rho_{fis}$ at physiological and atmospheric oxygen. In all cases- diamonds (◊): physiological cell density ($\rho_{fis}$) at physiological ambient oxygen concentrations, asterisk (\*): $\rho_{fis}$ in ambient atmospheric oxygen, circles (o): $\rho_{fis}/10$ at atmospheric $O_2$, triangles (Δ): $\rho_{fis}/100$ with $O_2=0.03$ moles/m³, pluses (+):$\rho_{fis}/100$ with atmospheric $O_2$ (0.2 moles/m³).*

DISCUSSION

This paper establishes a theoretical framework for estimating cellular metabolic rate in-vitro. Using finite element methods and models of 3D spheriods, the analysis shows that even in the absence of vascularisation, cells and tissues can maintain quarter power scaling in culture under specific conditions. The main experimental variables which determine whether a construct lies in the physiological scaling window are: its size, the cell density and the ambient or surface oxygen concentration. Physiological or "metabolically-supported functional scaling" [14] can be maintained within a "working window" when between 5 and 60% of the cells in the construct volume are exposed to oxygen levels below $k_m$. This corresponds to a Thiele modulus of between 8 and 80 and challenges the long held assumption that tissue oxygen gradients should be minimized and $\phi^2$ should be close to 1 [26].

The fundamental assumptions behind the models are i) there are no convection driven distribution networks inside the constructs, and oxygen transport is driven by Michaelis Menten reaction kinetics and diffusion; ii) the oxygen levels at the surface of the constructs are constant, hence the media is well-mixed and renewed continuously, as may be the case in fluidic systems; iii) the limiting values of CMR and $k_m$ are constant, independent of construct size.

Although specific values of CMR and $k_m$ were used for the computation, the results can be generalized for the fairly restricted ranges of CMR and $k_m$ reported in the literature. The limiting CMR determines the absolute maximum value of the curves in Figure 2 and 3, in accordance with Eq. 6. It corresponds to the maximum consumption rate of a single isolated mammalian cell in the presence of high oxygen levels, and represents the metabolic rate to which all mammalian cells converge when cultured in standard in-vitro conditions [7]. On the bases of the considerations outlined here, it is simple to demonstrate that cells isolated from an organism and cultured in low density monolayers in media exposed to atmospheric oxygen have a higher mass specific metabolic rate (CMR) than in-vivo. On the other hand, when ambient oxygen concentrations are lower, as in some of the cases shown in Figure 3, the maximum CMR is reduced due to lower oxygen availability (equation A9 in the Appendix.). The mass at which the slope in Figures 2 and 3 transitions from zero to negative values does depend on the Michaelis Menten constant, but the dependence is only significant when $k_m \sim c_o$, an experimentally unlikely case (data not shown).

A number of reports have confirmed that the cellular oxygen consumption rate in 3D constructs is lower than in monolayers [27–29]. In fact, probably the simplest way to establish whether tissues in-vitro are obeying quarter power scaling is to measure tissue oxygen consumption rates as a function of construct size keeping cell density and ambient oxygen constant. Since $M \propto R^3$, should quarter power scaling hold, a 25% increase in construct dimensions will result in an ≈15% decrease in CMR. On the other hand, if the rates remain constant, then the tissue does not obey physiological allometric scaling laws and the construct size should be increased incrementally until the CMR just begins to fall.

Very few cells in the body are ever exposed to 0.2 moles/m³ $O_2$ under normal physiological conditions and most tissues thrive at between ≈0.1 and 0.01 moles/m³ [20]. There is indeed a growing awareness among scientists that current in-vitro methods do not mimic the oxygen levels observed in-vivo, limiting the predictive value of cell cultures particularly as regards metabolic functions (14, 25). The analysis presented here suggests that the development of physiologically relevant in-vitro models with translational value requires a change in experimental paradigms as well as the development of supporting technology for monitoring and regulating oxygen. Cells should be cultured in 3D, in larger scaffolds and higher densities than previously thought acceptable and gradually coerced to re-adapt to lower oxygen concentrations and concentration gradients either through proliferation and migration [28,30] or a controlled reduction in ambient oxygen [31]. The results have important implications for the design of more predictive and physiologically relevant fluidic devices and organ-on-a-chip systems.

APPENDIX

Here the CMR and its relationship with mass in a non-vascularised cell-filled spherical construct for Michaelis-Menten reaction kinetics is derived from first principles in conditions of i) $c>>k_m$ and ii) $c<<k_m$.

The MM equation for oxygen consumption is:

$$\frac{dc}{dt} = \frac{V_{max} c}{k_m + c} \qquad (A1)$$

The parameters and their units are given in the main text. When $c>>k_m$, oxygen consumption is zero order, that is constant and independent of $c$:

$$\frac{dc}{dt} = V_{max} \qquad (A2)$$

When $c<<k_m$, the reaction rate depends on $c$ and is first order.

$$\frac{dc}{dt} = \frac{V_{max}}{k_m} c = Kc \qquad (A3)$$

Combining the reaction equations with diffusion of oxygen within the volume (in spherical coordinates and considering spherical symmetry) gives respectively:

$$\frac{\partial c}{\partial t} = \frac{D}{r^2} \frac{\partial}{\partial r}\left(r^2 \frac{\partial c}{\partial r}\right) - V_{max} \qquad \text{zero order}$$

$$\frac{\partial c}{\partial t} = \frac{D}{r^2} \frac{\partial}{\partial r}\left(r^2 \frac{\partial c}{\partial r}\right) - \frac{V_{max}}{k_m} c \qquad \text{first order}$$

For zero order consumption, the CMR can be easily derived for steady state conditions (Eq 5 main text) considering a constant oxygen concentration at the surface of a construct with cell density ρ.

$$CMR = \frac{V_{max}}{\rho} \qquad (A4)$$

There is no dependence on the size of the construct and it does not follow physiological metabolic scaling. The CMR is the classic expression for the so-called oxygen consumption rate or OCR in-vitro.

In the case of first order kinetics, for a sphere with radius $R$ and a constant concentration of $c=c_o$ at its surface, the equation can be solved for stationary conditions, and expressed in terms of Thiele's modulus ($\phi^2 = R^2 V_{max} / k_m D$) giving:

$$c = c_o \frac{R}{r} \frac{\sinh\left(r\sqrt{V_{max}/k_m D}\right)}{\sinh\left(R\sqrt{V_{max}/k_m D}\right)} = c_o \frac{R}{r} \frac{\sinh\left(\frac{r}{R}\phi\right)}{\sinh(\phi)} \quad (A5)$$

Again, the overall metabolic rate, MR in the sphere is the total inward flux at its surface multiplied by its surface area.

$$MR = D\frac{dc}{dr}\bigg|_R 4\pi R^2 \quad (A6)$$

Thus differentiating Eq. A1 and substituting r for R in Eq. 6 gives :

$$MR = 4\pi c_o RD \left(\frac{\phi}{\tanh(\phi)} - 1\right) \quad (A7)$$

The average CMR is the MR divided by the total number of cells in the sphere ($4\pi R^3 \rho/3$),

$$CMR = \frac{3c_o D}{\rho R^2}\left(\frac{\phi}{\tanh(\phi)} - 1\right) \quad (A8)$$

Given that $k_m$ and $V_{max}$ are considered scale invariant, Thiele's modulus depends on R. For spheres with small radii, i.e. $\phi^2 \ll 1$, $\tanh(\phi)$ can be expressed as the first two terms of the series expansion.

$$CMR = \frac{3c_o D}{\rho R^2}\left(\frac{\phi}{\phi - \frac{\phi^3}{3}} - 1\right) = \frac{c_o}{\rho}\frac{V_{max}}{k_m} \quad (A9)$$

The CMR does not depend on the size of the construct and the allometric exponent b=0.

On the other hand, for large tissue constructs, $\phi^2 \gg 1$ and $\tanh(\phi) \to 1$. In this case the CMR is:

$$CMR = \frac{3c_o}{\rho R}\sqrt{\frac{V_{max} D}{k_m}} \quad (A10)$$

which can also be expressed in terms of the mass of the construct (water density $\Omega$ × volume) to highlight its non-physiological b=-1/3 allometric scaling.

$$CMR = \left[\frac{3c_o}{\rho\sqrt[3]{3/4\pi\Omega}}\sqrt{\frac{V_{max} D}{k_m}}\right] M^{-1/3} \quad (A11)$$

Figure A1 shows a log-log graph of CMR against the mass of a cell construct with first order consumption using the values of cell density ρ, $V_{max}$ and $k_m$ listed in Table 1 and with $c_o$ = 0.2 moles/m³ (atmospheric dissolved $O_2$). In between the two extremes of $\phi^2$ which result in b=0 and b=- 1/3 respectively, for each of the curves reported in Figure A1 lies a single value of mass and Thiele's modulus where b=- ¼. (Figure A2 and A3).

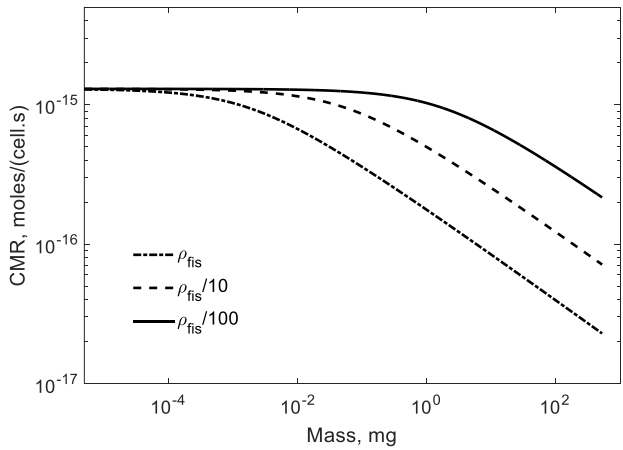

A1

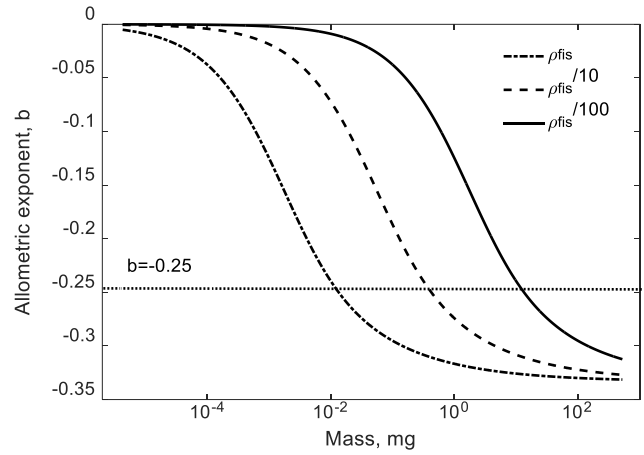

A2

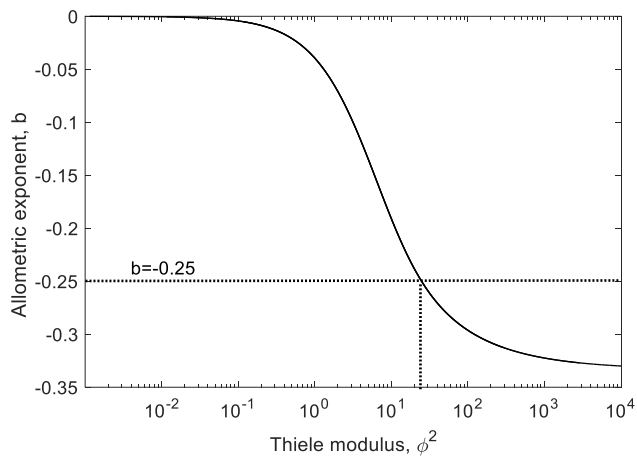

A3

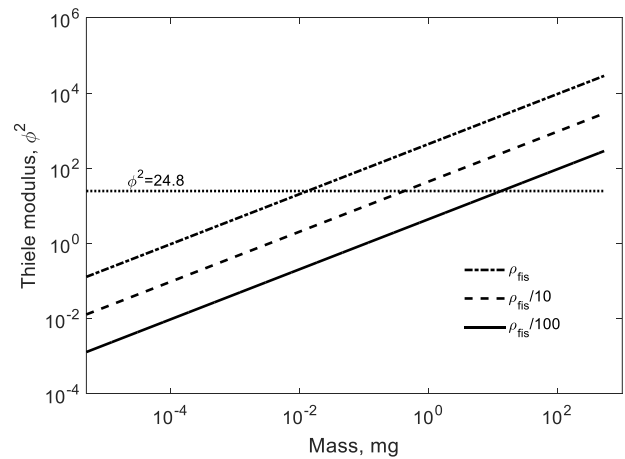

A4

***Figures A1 to A4. Analysis of quarter power scaling in constructs with first order oxygen consumption.*** *Dotted and dashed line (.-) : physiological density $\rho_{fis}$= 5.14×10¹⁴ cells/m³, dashed line (--): $\rho_{fis}$ /10, full line: $\rho_{fis}$/100. A1) Log scale plot of CMR versus mass. A2) Numerical derivative of A2 (i.e. the allometric exponent b) and line corresponding to b=-1/4. A3) Allometric exponent b (from A2) versus Thiele modulus highlighting the unique value of $\phi^2$ for which b=-1/4. The three curves for different values of $\rho$ are coincident. A4) Thiele modulus versus mass. The dotted line is the value of $\phi^2$ which corresponds to b=-1/4 (from Figure A3).*

To solve for the values of M and $\phi^2$ which correspond to b=- ¼ , equation A8 can be expressed in terms of construct mass:

$$CMR = \frac{3Dc_o}{\rho \left(\frac{3}{4\pi\Omega}\right)^{2/3} M^{2/3}} \left[ \frac{\left(\frac{3}{4\pi\Omega}\right)^{1/3} M^{1/3} \sqrt{\frac{V_{max} D}{k_m}}}{\tanh\left(\left(\frac{3}{4\pi\Omega}\right)^{1/3} M^{1/3} \sqrt{\frac{V_{max} D}{k_m}}\right)} - 1 \right] \quad (A12)$$

and the following identity solved numerically in Matlab (using the in-built optimization functions).

$$\frac{M}{CMR} \frac{d(CMR)}{dM} + 0.25 = 0 \quad (A13)$$

The analysis shows that the value of M which respects quarter power scaling depends on the variables ρ, D, $V_{max}$, $k_m$ but not on $c_o$. However, for all variables investigated and reported in Table 1, $\phi^2$ =24.808 when b=- 1/4. This can be confirmed graphically (Figure A3 and 4).